\documentclass[preprint,12pt,number]{elsarticle}

\usepackage[T1]{fontenc}
\usepackage{amsmath,amssymb}
\usepackage{graphicx}
\usepackage{booktabs}
\usepackage{lineno}            % elsarticle review mode uses line numbers
\usepackage{url}
\usepackage{hyperref}
\hypersetup{hidelinks}
\journal{Acta Astronautica}

\begin{document}

\begin{frontmatter}

% --- Title ---
\title{Kardashev's Conundrum: Statistical
Falsification of the Standard One-Percent
Kardashev Growth Model, the Non-Observation of
Type~II Technosignatures, and the
Kardashev--Sagan--Nakamoto Resolution}

% --- Author and affiliations ---
\author[iate,wsu]{S.~Gurovich\corref{cor1}}
\ead{sgurovich@unc.edu.ar}
\cortext[cor1]{Corresponding author.}
\nonumnote{\raggedright Accepted for publication
in Acta Astronautica, November 2026.
\copyright~2026. This manuscript version is made
available under the CC BY-NC-ND 4.0 licence,
\url{https://creativecommons.org/licenses/by-nc-nd/4.0/}.}

\address[iate]{Instituto de Astronom\'ia Te\'orica y Experimental
(IATE--OAC--UNC--CONICET), Laprida 854, X5000BGR, C\'ordoba, Argentina}

\address[wsu]{Western Sydney University, Kingswood campus, NSW, Australia}

% --- Abstract ---
\begin{abstract}
We test the standard Kardashev conjecture -- that
a technologically advanced civilisation's energy
production grows at a fixed one percent per year
-- against six decades of global energy-production
data (1965--2024), with consequences for the
search for extraterrestrial intelligence.
Empirically, the specific one-percent
extrapolation is not supported by the terrestrial
energy record: Markov Chain Monte Carlo inference
returns a posterior growth rate
$r=2.01\%\pm0.03\%~\mathrm{yr^{-1}}$. Neither of
the two simplest statistically adequate models (a
linear form and a free-rate exponential, between
which model selection shows no significant
preference) yields a physically
coherent path to the energy-scaling Type~II
threshold, the solar luminosity: the linear form
takes of order $10^5$ Hubble times to reach it;
the exponential, sustained on Earth, would make
the planet uninhabitable through waste heat long
before. This is Kardashev's Conundrum. Whatever
its growth law, a planet-bound trajectory crosses
the waste-heat ceiling of habitability before the
Type~II threshold, so under stated assumptions
Dyson spheres, if they exist, are the product of
relocating energy dissipation off the home planet:
a candidate explanation for the null results of
Dyson-sphere surveys. Conceptually, the Kardashev
state variable, power, is dimensionally
incomplete: it captures energy throughput but not
the information richness of its use. We therefore
propose the Kardashev--Sagan--Nakamoto
renormalisation, dividing energy production by the
Bitcoin proof-of-work hashrate, an observational
proxy for one measurable class of global
computation, to define $B=P/H$ in
$\mathrm{J\,Hash^{-1}}$ -- the KarNak unit -- a
dimensionally complete information--energy
variable motivated by the Landauer limit, on which
the Type~II concept is preserved.
\end{abstract}

% --- Keywords ---

\end{frontmatter}

% In review mode, uncomment to number lines:
% \linenumbers

% ================================================
%  The sections of the paper (body.tex).
% ================================================

\section{Introduction}
\label{sec:intro}

In a landmark 1964 paper, \citet{Kardashev1964} proposed a universal scale for
classifying the technological advancement of civilisations by their total rate of
energy consumption. A Type~I civilisation harnesses all the energy available on its
home planet -- \citet{Kardashev1964} estimated Earth's total expenditure at the time
at $4\times10^{12}$~W -- a Type~II civilisation consumes the entire energy output of
its host star ($L_\odot \approx 3.828\times10^{26}$~W), and a Type~III civilisation
commands the output of its host galaxy. Adopting a conservative growth rate of one
percent per year, \citet{Kardashev1964} calculated that humanity would reach Type~II
status in approximately $3{,}200$ years and Type~III in approximately $5{,}800$ years.

The scale's enduring scientific importance is
observational: it defines the targets of the
search for extraterrestrial intelligence (SETI).
The canonical realisation of a Type~II
civilisation is a Dyson sphere or swarm -- an
artificial structure intercepting a star's full
luminosity \citep{Dyson1960} -- whose reprocessed
waste heat would appear as an anomalous
mid-infrared excess. This prediction has driven a
sustained observational programme: the whole-sky
upper limit on Dyson spheres from Infrared
Astronomical Satellite (IRAS) data
\citep{Carrigan2009}, the $\hat{G}$ infrared
search for civilisations with large energy
supplies \citep{Wright2014}, and recent candidate
searches combining photometry from Gaia Data
Release 3, the Two Micron All Sky Survey (2MASS),
and the Wide-field Infrared Survey Explorer (WISE)
\citep{Suazo2024}. To date these surveys have
yielded no confirmed Type~II technosignature. The
theoretical and observational foundations of the
field are reviewed comprehensively by
\citet{LingamLoeb2021} and, most recently, in the
first dedicated SETI textbook of
\citet{Wright2026}.

The Kardashev scale has been extended and
critiqued in several directions. \citet{Gray2020}
refined its definitions and resolved long-standing
ambiguities; \citet{Barrow1998} proposed a
complementary ``inward'' scale, ranking a
civilisation's mastery of progressively smaller
length scales rather than larger energy budgets.
Most relevant to the present work,
\citet{Ivanov2020} argued that contemporary
searches are calibrated to detect only
civilisations that share our own
``more-is-better'' energy philosophy, and
suggested that the persistent failure of
Dyson-sphere searches may reflect the inadequacy
of that assumption rather than the absence of
civilisations. This paper supplies the
quantitative counterpart to that qualitative
argument: a statistical falsification, against six
decades of data, of the fixed one-percent growth
rate that Kardashev extrapolated to Type~II, and a
physical argument, which holds whatever the growth
law, that planet-bound growth reaches the
waste-heat ceiling of habitability before the
energy-scaling Type~II target.

Despite its influence, the Kardashev scale carries a limitation that has received
little quantitative attention. \citet{Kardashev1964}, \citet{Sagan1973}, and
\citet{ShklovskiiSagan1966} all noted that an advanced civilisation must be
characterised not only by its energy throughput but by the information richness of what
it does with that energy; \citet{Sagan1973} proposed that a civilisation's state should
reflect its capacity to secure and transmit vast quantities of information, and
suggested a binary-tree data structure as a measure of informational complexity.
Neither author, however, formalised this requirement as a quantitative variable within
the framework. A civilisation that wastes energy inefficiently is rated identically, on
the Kardashev scale, to one that channels the same power into sophisticated
computation. The Kardashev state variable -- power, in watts -- is therefore
dimensionally incomplete as a measure of civilisational advancement: it captures the
quantity of energy consumed but not the quality of its use.

A second limitation emerges from direct comparison
with data. When the one-percent exponential model
of \citet{Kardashev1964} is compared against six
decades of observed total global energy production
\citep{Ritchie2022} -- used here as a valid
upper-limit proxy for consumption, since
production necessarily equals or exceeds
consumption at the civilisational scale and the
two track each other closely over decadal
baselines -- the model diverges substantially from
the data. A linear model fits the same data
well, yet when extrapolated to the Type~II
threshold it predicts a transition timescale of
order $10^5$ Hubble times: a physical
\emph{reductio ad absurdum}. A free-rate
exponential fits the data equally well, but growth
at its fitted rate, sustained on Earth, would
render the planet uninhabitable through waste heat
long before the Type~II threshold is reached
\citep{BalbiLingam2025}. We term this conflict
\emph{Kardashev's Conundrum}: neither of the two
simplest \emph{statistically adequate} models of
the observed energy record (a linear form and a
free-rate exponential, between which model
selection shows no significant preference) yields
a physically coherent path to
the energy-scaling Type~II threshold.

Two results follow. First, the Conundrum has an immediate observational payoff for
SETI. If the energy-scaling Type~II threshold is not
reached on planet-bound trajectories, then
Type~II energy-harvesting technosignatures are
unlikely to arise from civilisations that
follow them. We develop
this in Section~\ref{sec:technosignatures} as a quantitative entry in the ``logistic''
category of Fermi-paradox solutions of \citet{Cirkovic2018}, where it draws on the waste-heat constraint of
\citet{BalbiLingam2025} and joins the
luminosity-limit constraint of
\citet{HaqqMisra2025b} on sustained
technological growth. Second, the Conundrum motivates a
constructive repair. To resolve it we propose the Kardashev--Sagan--Nakamoto (KSN)
model, constructed by renormalising the energy-production ordinate by the annual average
Bitcoin network hashrate $H(t)$. The resulting state variable $B = P/H$, in units of
$\mathrm{J\,Hash^{-1}}$ -- which we term the KarNak unit -- connects energy to
irreversible computational work via the Landauer limit \citep{Landauer1961} and fulfils
the information-richness requirement identified but never quantified by
\citet{Kardashev1964} and \citet{Sagan1973}. The Bitcoin hashrate is used here strictly as an
observational proxy for one measurable class of
global computation: proof-of-work is one
specialised, economically motivated workload, not
the total computation of civilisation, but this
hashrate is the dominant and longest-running
publicly auditable, unforgeable, and continuously
updated global record of any class of
computation. Once the denominator is chosen, the
renormalisation introduces no free parameter:
$B$ is the ratio of two quantities measured
from separate records.

The notion that civilisational advancement is
measured by the efficiency with which energy is
converted into organised information is an ancient
one. The Antikythera mechanism, recovered from a
first-century BC shipwreck, is widely recognised
as the earliest known analogue computational
device \citep{Freeth2021}; built from bronze gears
of remarkable tolerance, it converted mechanical
energy into astronomical information with a
sophistication not matched in Europe for over a
millennium. From Babbage's difference engine
through Turing's theoretical foundations of
computation \citep{Swade2017} to the
application-specific integrated circuits (ASICs)
that today secure the Bitcoin network, humanity
has followed a continuous trajectory of reducing
the energy cost per unit of computational work.
The KSN variable $B(t)$ follows this trajectory
for one class of computation, although not in
isolation: over 2009--2024 it spans fourteen
orders of magnitude, reflecting both the improving
efficiency of the hardware from the central
processing unit (CPU) era to modern ASICs and the
growing share of global energy devoted to the
network.

This paper is organised as follows.
Section~\ref{sec:analysis} describes the data
(Section~\ref{sec:data}) and presents the
statistical core: the assessment of the standard
Kardashev model (Section~\ref{sec:standard}), the
linear model (Section~\ref{sec:linear}), Bayesian
model comparison together with the full statement
of the Conundrum (Section~\ref{sec:mcmc}), the KSN
renormalisation and its physical motivation
(Section~\ref{sec:ksn}), civilisational timescales
on the KSN scale (Section~\ref{sec:timescales}),
and the independence-property objection to
exponential growth
(Section~\ref{sec:independence}).
Section~\ref{sec:discussion} opens with the
observational consequence for Type~II
technosignatures
(Section~\ref{sec:technosignatures}) before
turning to falsifiability
(Section~\ref{sec:falsifiability}), limitations
(Section~\ref{sec:limitations}), the novel reading
of Kardashev's two growth rates
(Section~\ref{sec:growthrates}), and the
connection to the Drake equation
(Section~\ref{sec:drake}).
Section~\ref{sec:conclusions} presents our
conclusions.

% =====================================================================
\section{Kardashev's Conundrum and the Kardashev--Sagan--Nakamoto Model}
\label{sec:analysis}

We now present our central analysis. We show that
the total global energy-production data reject the
standard Kardashev one-percent exponential model,
and that the two simplest descriptions that do fit
the data, a linear Ordinary Least-Squares (OLS)
model and a free-rate exponential, with no
significant preference between them, fail in
complementary ways when extrapolated to the
Type~II threshold. This is Kardashev's Conundrum.
We then propose the Kardashev--Sagan--Nakamoto
(KSN) renormalisation, which divides energy
production by the measured rate of one class of
computation.

\subsection{Data}
\label{sec:data}

The energy-production data $D$ consist of the
world total primary energy consumption in
terawatt-hours per year ($\mathrm{TWh\,yr^{-1}}$),
drawn from the Our World in Data (OWID) energy
dataset \citep{Ritchie2022}, which integrates the
Energy Institute (EI) Statistical Review of World
Energy \citep{EI2025}, the US Energy Information
Administration (EIA) international energy series
\citep{EIA2025} and, for electricity generation,
the Ember yearly electricity data
\citep{Ember2026}; see \citet{Smil2017} for a
comprehensive historical analysis of the
industrial-era energy trajectory. We use $N=60$
annual data points spanning 1965--2024, from the
release of 27 April 2026. In that release the
column \texttt{primary\_energy\_consumption}
follows the input-equivalent (substitution)
convention, in which electricity
generated from nuclear, hydro, wind and solar
sources is counted as the fossil fuel that would
have been needed to generate it, at an assumed
conversion efficiency of 36 percent until 2000,
rising to about 41 percent by 2024. OWID has
since followed the EI in measuring primary energy
as total energy supply \citep{Ritchie2026}, which
counts hydro, wind and solar electricity as
generated and nuclear electricity as the heat
released in the reactor. In the OWID release of
10 September 2026 the world totals are lower than
those used here, by 4 percent for 1965 and by
7 percent for 2024, the first and last years of
the record.
Section~\ref{sec:limitations} repeats the linear
and exponential fits on two other series: the
total energy supply of the release of
10 September 2026, and the series used here with
the electricity from all four sources counted as
generated.
On all three series the one-percent
rate is excluded and neither the linear nor the
exponential form is significantly preferred by
model selection, although how strongly one form is
favoured over the other differs from one series to
another.

All energy values are converted to SI watts
($\mathrm{J\,s^{-1}}$) assuming uniform power
delivery over each calendar year via
\begin{equation}
\label{eq:conv}
P\,[\mathrm{W}] = E\,[\mathrm{TWh\,yr^{-1}}] \times
\frac{10^{12}\,[\mathrm{Wh\,TWh^{-1}}]
      \times 3600\,[\mathrm{J\,Wh^{-1}}]}
     {365.25 \times 24 \times 3600
      \,[\mathrm{s\,yr^{-1}}]},
\end{equation}
where the numerator converts terawatt-hours to
joules ($1~\mathrm{TWh}=3.6\times10^{15}$~J) and
the denominator is the number of seconds in one
Julian year. The factors of $3600$ cancel,
reducing Equation~(\ref{eq:conv}) to
$P = E\times10^{12}/(365.25\times24)$~W. This
yields $P(1965)=4.95$~TW, in close agreement with
the \citet{Kardashev1964} baseline of $4$~TW, and
$P(2024)=20.2$~TW. We set the origin of the time
coordinate at $t_0=1964$, the year of the
Kardashev conjecture \citep{Kardashev1964}, so
that $t=\mathrm{year}-1964$ throughout.

The Bitcoin network hashrate data $H(t)$ are
yearly averages in $\mathrm{Hash\,s^{-1}}$
for 2009--2024. These span the full history of
Bitcoin mining from the CPU era
($\sim4.5~\mathrm{MH\,s^{-1}}$ in 2009) through
the ASIC era
($\sim6.3\times10^{20}~\mathrm{H\,s^{-1}}$ in
2024), covering 16 annual data points and 14
orders of magnitude in hashrate. Every yearly
average is computed in the same way, from the
blocks mined in that year. The difficulty of a
block is a dimensionless number that the
protocol recalculates every 2016 blocks and
that equals one at the easiest setting it
allows. The number of evaluations of the hash
function needed to find a block is on average
$2^{32}$ times its difficulty, so
the average hashrate over a calendar year is
$2^{32}$ times the summed difficulty of the
blocks mined in that year, divided by the number
of seconds in the year. The hash function and the
definition of the difficulty were the same
throughout 2009--2024, so the series is a
single homogeneous measurement. The daily values
of this quantity are published in the Coin
Metrics community data set
\citep{CoinMetrics2026}, and $H(t)$ is their
mean over each calendar year.

\paragraph{Computation and the use of an
artificial intelligence (AI) tool}
The figures and the quantitative results of this
paper are produced by Python scripts in a public
repository (see the Data availability section).
The scripts read the copies of the data kept
there and use fixed random seeds, so that every
result can be reproduced by running them, without
any AI tool. They were written or revised, and
tested, with the assistance of Claude
(Anthropic), in successive versions of its Opus
and Fable models, the latest being
\mbox{Opus~5.5} and \mbox{Fable~5.1}. The same
tool assisted with statistical and dimensional
checks, and with the derivation and checking of
the number of bits erased per hash
(Section~\ref{sec:ksn}) and of the timescale
estimate of Section~\ref{sec:timescales}. The
author ran the scripts himself, on his own
computer and on the S\'ersic cluster (see the
Acknowledgements), and reviewed the derivations
and the results.

\subsection{The Standard Kardashev Model and its Statistical Assessment}
\label{sec:standard}

\citet{Kardashev1964} proposed that the energy
production of a technologically advanced
civilisation increases at a constant fractional
rate per year, giving a geometric series
\begin{equation}
\label{eq:kardashev}
P_K(t) = P_0 (1+x)^t \approx P_0\, e^{xt},
\qquad x = 0.01,
\end{equation}
where $P_0$ is the energy production at $t=0$
(1964) and $x=0.01$ corresponds to the stated
one-percent per-year growth rate. Using this
model, \citet{Kardashev1964} calculated that a
Type~II civilisation -- defined as one harnessing
the total bolometric luminosity of its host star,
$L_\odot=3.828\times10^{26}$~W \citep{Mamajek2015}
-- would be reached in approximately $3{,}200$
years from 1964, corresponding to approximately
year 5200~CE. \citet{Sagan1973} obtained a similar
value using a slightly revised Type~II threshold
of $\sim10^{26}$~W.

\begin{figure}[htbp]
\centering
\includegraphics[width=\linewidth]{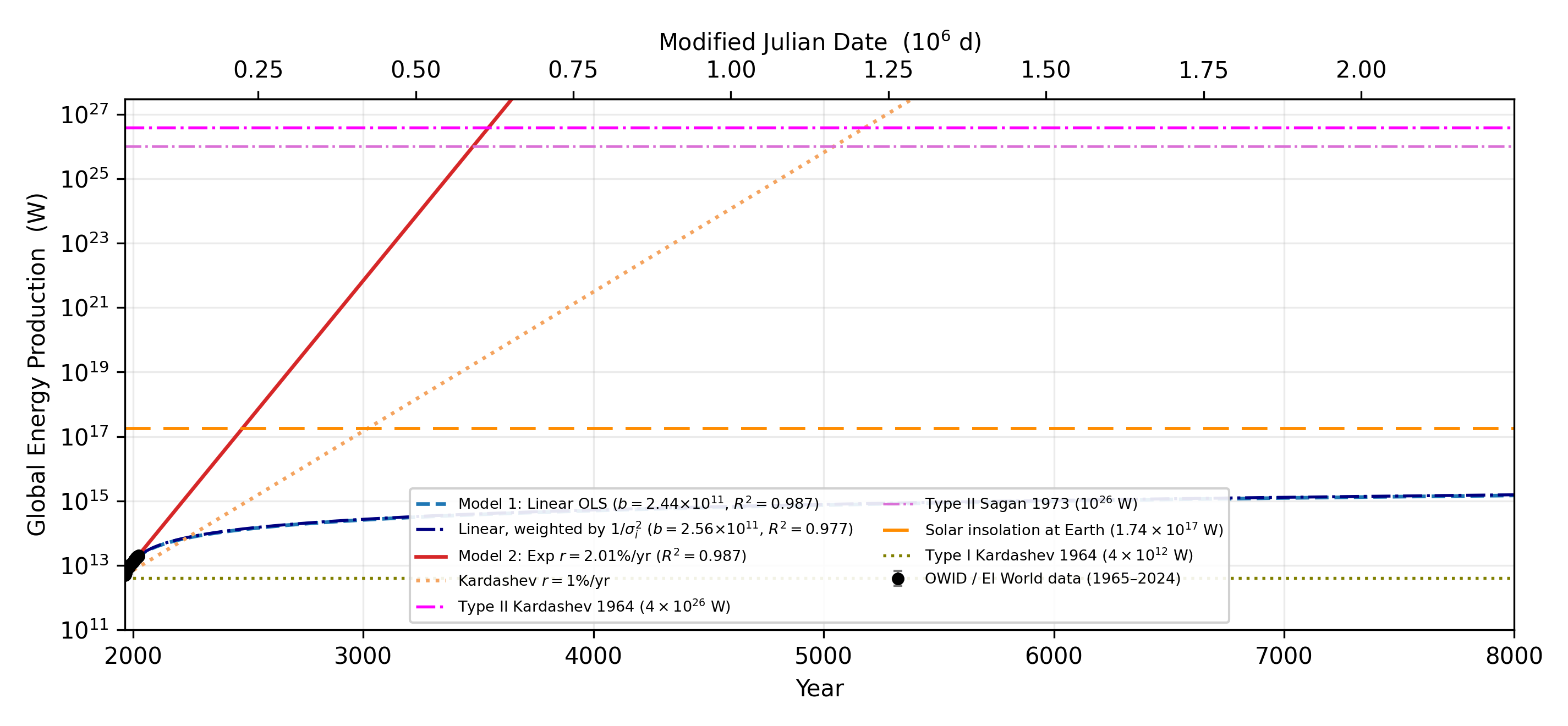}
% --- error bars are the adopted heteroscedastic model (Sec. 3.3); both the
%     unweighted (OLS) and $1/\sigma_i^2$-weighted linear fits are shown. The
%     figure is produced by make_errorbar_figures.py on the OWID data. ---
\caption{Global energy production 1965--2024
(OWID) in watts (filled circles), with the
unweighted best-fit linear model (Model~1, blue
dashed; $b=2.44\times10^{11}$~W~yr$^{-1}$,
$R^2=0.987$), the $1/\sigma_i^2$-weighted linear
fit (navy dash-dot;
$b=2.56\times10^{11}$~W~yr$^{-1}$, $R^2=0.977$),
the free-rate exponential fit (Model~2, red
solid; $r=2.01\%~\mathrm{yr^{-1}}$, $R^2=0.987$),
the standard Kardashev one-percent exponential
(salmon dotted), and the Kardashev Type~I,
Type~II \citep{Kardashev1964,Sagan1973} and
solar-insolation reference levels (horizontal
lines). The one-percent model diverges
visibly from the data over the full six-decade
baseline. The OWID series reports point estimates
without published per-year uncertainties; we
therefore assign to each yearly value an assumed
uncertainty that varies with the year
(Section~\ref{sec:limitations}), a fraction of
the value that declines from $\sim5$--$10\%$
before 1980 to $\sim1$--$2\%$ after 2000,
reflecting improving reporting coverage. On this
logarithmic ordinate the corresponding error bars
are smaller than the plotting symbols. Refitting
under $1/\sigma_i^2$ weighting leaves every
qualitative conclusion unchanged: the one-percent
Kardashev model remains decisively rejected, and
the Widely Applicable Information Criterion (WAIC)
still shows no significant preference between the
linear and free-rate exponential models
(Section~\ref{sec:mcmc}).}
\label{fig:energy_loglog}
\end{figure}

Figure~\ref{fig:energy_loglog} shows the Kardashev model (Eq.~\ref{eq:kardashev})
alongside the OWID data, together with the Type~I and Type~II civilisation limits and
the solar insolation at Earth ($F_\oplus\approx1.74\times10^{17}$~W). The one-percent
Kardashev model diverges visibly from the data over the full six-decade baseline.

We quantify this divergence statistically. The
likelihood of the data given the Kardashev model
with a point prior on $x=0.01$ is
\begin{equation}
\label{eq:like}
\mathcal{L}(D \mid \theta_K, M_K) = \prod_{i=1}^{N}
\frac{1}{\sqrt{2\pi\sigma^2}}\,
\exp\!\left[ -\frac{\left(P_i - P_0
(1.01)^{t_i}\right)^2}{2\sigma^2} \right],
\end{equation}
where $\sigma^2$ is estimated from the residuals.
The residual sum of squares for the Kardashev
model exceeds that of the linear model by more
than two orders of magnitude, constituting
overwhelming evidence against the one-percent
prior. We therefore conclude that the standard
Kardashev exponential model with $x=0.01$ is
statistically untenable: the specific one-percent
extrapolation is not supported by the 1965--2024
terrestrial energy record.

\subsection{The Linear Model: an Excellent Fit with a Physical Reductio}
\label{sec:linear}

\begin{figure}[htbp]
\centering
\includegraphics[width=\linewidth]{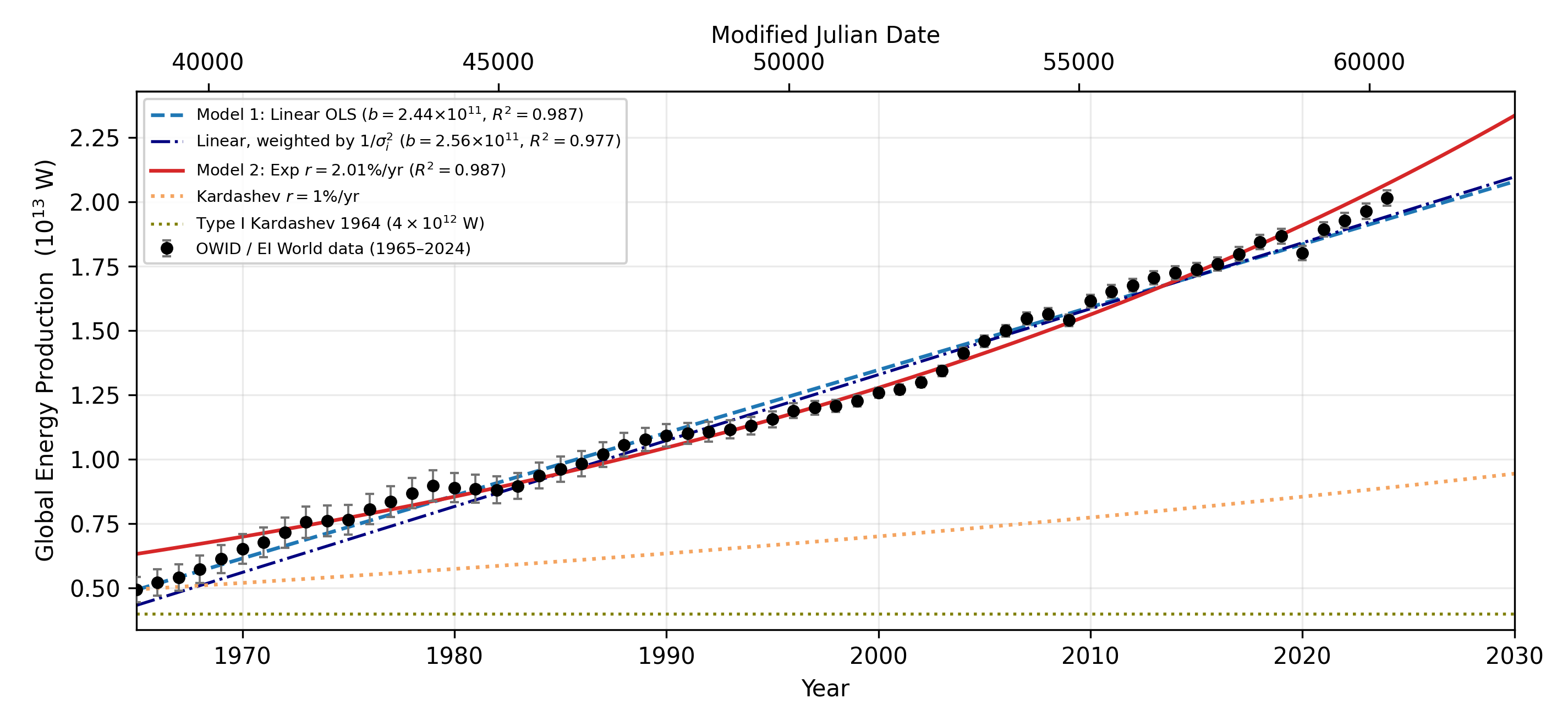}
\caption{Re-scaled version of the OWID global
primary-energy data (1965--2024) on a linear
vertical axis, in units of $10^{13}$~W, with the
unweighted linear OLS fit (blue dashed,
$b=2.44\times10^{11}$~W~yr$^{-1}$, $R^2=0.987$),
the $1/\sigma_i^2$-weighted linear fit (navy
dash-dot, $b=2.56\times10^{11}$~W~yr$^{-1}$,
$R^2=0.977$), and the free-rate exponential fit
($r=2.01\%~\mathrm{yr^{-1}}$; red solid,
$R^2=0.987$). The two linear fits and the
exponential are visually near-identical over the
six-decade baseline, consistent with the WAIC result
that neither functional form is significantly
preferred on the observed data. Vertical
error bars show the adopted per-year uncertainties
(Section~\ref{sec:limitations}): comparable to or
smaller than the marker size for post-2000 data
and growing to $\sim5$--$10\%$ for the earliest
points. The Kardashev one-percent exponential
(salmon dotted) diverges dramatically downward
from the data, falsifying the conservative
assumption widely adopted in the subsequent
literature. The weighting raises the linear slope
by $\approx5\%$ (from $b=2.44\times10^{11}$ to
$2.56\times10^{11}$~W~yr$^{-1}$), as the
higher-uncertainty early points are down-weighted,
but leaves the WAIC comparison of
Section~\ref{sec:mcmc} inconclusive and does not
alter the order-of-magnitude Type~II timescale;
the Shapiro--Wilk non-Gaussianity
(Figure~\ref{fig:deltaP}) is driven by the
well-measured 2009 and 2020 outliers and is
insensitive to the early-decade weighting. The
Kardashev Type~I threshold from 1964
($4\times10^{12}$~W) lies well below the full
dataset: Earth has long since surpassed
Kardashev's original Type~I energy benchmark.}
\label{fig:energy_linear}
\end{figure}

We next consider a linear OLS model
\begin{equation}
\label{eq:linear}
P_L(t) = a + b\,t,
\end{equation}
with parameters $a$ and $b$ estimated by
minimising the sum of squared residuals. Following
Occam's razor \citep{Blumer1987}, this
two-parameter model is the simplest physically
plausible alternative. The OLS solution gives
$b=(2.44\pm0.04)\times10^{11}~\mathrm{W\,yr^{-1}}$
and coefficient of determination $R^2=0.987$, with
a root-mean-square residual of $0.48$~TW, 4\% of
the mean power over the record. The fit is
excellent over the six-decade baseline of
the data (Figure~\ref{fig:energy_loglog}).
Figure~\ref{fig:energy_linear} shows the same data
and fits on a linear vertical axis, making the
remarkably low dispersion directly visible: the
linear and exponential models are visually
near-identical across 1965--2024, while the
Kardashev one-percent model diverges strikingly
downward from the data.

\begin{figure}[htbp]
\centering
\includegraphics[width=0.85\linewidth]{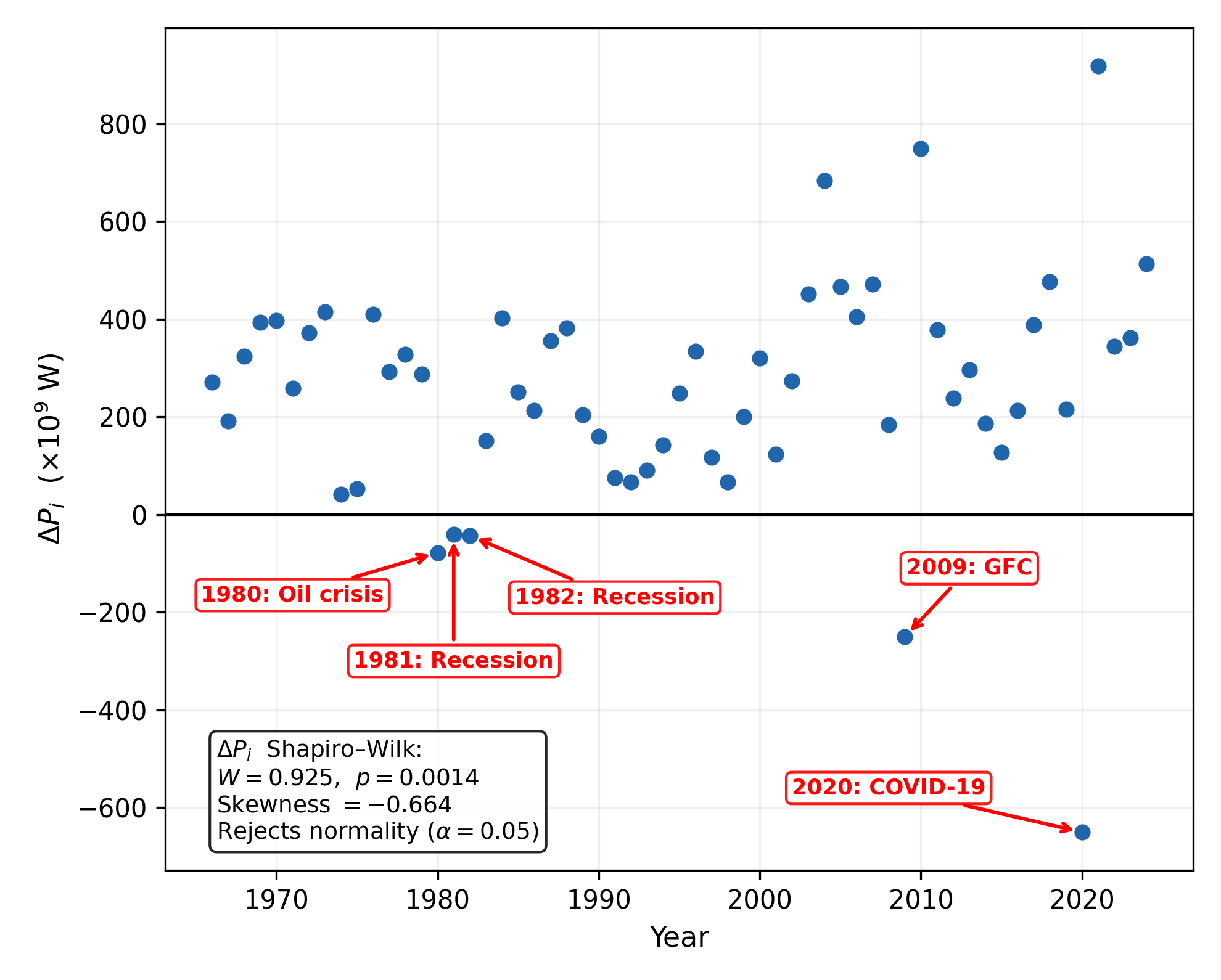}
\caption{Year-over-year increments
$\Delta P_i = P_i - P_{i-1}$ in global energy
production (1966--2024), in units of $10^9$~W.
The distribution is significantly non-Gaussian
(Shapiro--Wilk $W=0.925$, $p=0.0014$) and
negatively skewed (skewness $=-0.664$). The
largest negative excursions coincide with
identifiable historical shocks: the 1980--1982
oil crisis and recessions, the 2009 global
financial crisis (GFC), and the 2020 COVID-19
pandemic. These well-measured, recent-decade
outliers dominate the non-Gaussian signal and are
insensitive to the early-decade uncertainty
weighting discussed in
Section~\ref{sec:limitations}. Error bars are
deliberately not drawn on $\Delta P$: the adopted
per-year uncertainties
(Section~\ref{sec:limitations}) are dominated by
systematic effects -- accounting conventions,
reporting coverage, retroactive harmonisation --
that are strongly correlated between adjacent
years and therefore cancel in the difference
$P_i-P_{i-1}$. Independent propagation of the
full per-year uncertainty would overstate the
error on $\Delta P$ by a large factor (yielding
bars comparable to the increments themselves)
and is inappropriate here; the year-to-year
change in global energy is in fact constrained
far more tightly than the absolute level.}
\label{fig:deltaP}
\end{figure}

To assess the normality of the annual changes in
the record, we compute the year-over-year
differences $\Delta P_i = P_i - P_{i-1}$ and
apply the Shapiro--Wilk normality test
\citep{ShapiroWilk1965}. We obtain $W=0.925$,
$p=0.0014$: we reject the null hypothesis that
$\Delta P$ is drawn from a Normal distribution.
The distribution of $\Delta P$
(Figure~\ref{fig:deltaP}) exhibits significant
negative skewness (skewness $=-0.664$), driven by
identifiable exogenous shocks to global energy
production in 2009 (subprime financial crisis) and
2020 (COVID-19 pandemic). This non-Gaussian
structure shows that large negative shocks are
systematically more probable than Gaussian
increments would predict, and their occurrence is
traceable to specific historical events rather
than symmetric statistical fluctuations. It does
not by itself show that successive changes depend
on one another, a question taken up in
Section~\ref{sec:independence}.

A further structural objection to $(1+x)^t$ with
$x>0$ is that the model is sign-constrained: every
year must exhibit positive growth by construction.
The observed series contains identifiable years of
negative growth (e.g.\ 2009 and 2020) that the
model cannot accommodate at any parameter value --
a structural, not merely statistical,
falsification of the always-positive growth
assumption that is independent of the
non-Gaussianity argument above.

Notwithstanding its excellent in-sample fit, the
linear model fails catastrophically on
extrapolation. Solving $P_L(t^\ast)=L_\odot$
gives
\begin{equation}
\label{eq:textrap}
t^\ast = \frac{L_\odot - a}{b} \approx
1.6\times10^{15}~\mathrm{yr}
\approx 1.1\times10^{5}\,H_0^{-1},
\end{equation}
where $H_0^{-1}\approx14$~Gyr is the Hubble time,
the reciprocal of the Hubble constant
$H_0\approx70~\mathrm{km\,s^{-1}\,Mpc^{-1}}$.
Timescales obtained by extrapolating a fit so far
beyond the data are order-of-magnitude values:
their uncertainty is set by the choice of
functional form, not by the statistical
uncertainty of the fit (1.5 percent in $b$).
The implied Type~II timescale exceeds the
age of the Universe by five orders of magnitude.
This is not merely improbable -- it
is physically incoherent, since it vastly exceeds
the main-sequence lifetime of the Sun. The result
is the first half of Kardashev's Conundrum, which
Section~\ref{sec:mcmc} completes: the linear
model, a statistically adequate description of
the data, predicts a civilisational transition
timescale that is cosmologically nonsensical.

\subsection{Model Comparison via MCMC and WAIC}
\label{sec:mcmc}

To place the comparison on a rigorous Bayesian
footing, we fit an exponential model with the
growth rate $r$ as a free parameter,
\begin{equation}
\label{eq:exp}
P_E(t) = a_0\, e^{rt},
\end{equation}
using a Metropolis--Hastings Markov Chain Monte
Carlo (MCMC) sampler
\citep{Metropolis1953,Hastings1970} with a prior
$r\sim\mathcal{N}(0.01, 0.02^2)$ centred on the
Kardashev value and $60{,}000$ posterior samples
after a $15{,}000$-step burn-in. The burn-in phase
is discarded because the first iterations of the
chain are influenced by the arbitrary starting
point in parameter space rather than by the target
posterior; the Markov property of the sampler
ensures that the chain eventually forgets this
initial condition and converges to the stationary
distribution, from which the retained samples are
drawn. The posterior mean growth rate is
$r=2.01\%\pm0.03\%~\mathrm{yr^{-1}}$ (95\%
credible interval $[1.94\%, 2.08\%]$), with the
Kardashev value of $1\%~\mathrm{yr^{-1}}$ lying
well outside the 95\% credible interval. The
best-fit exponential model gives $R^2=0.987$ and a
root-mean-square residual of $0.49$~TW, nearly
identical to the linear model
($\Delta R^2 < 0.001$).

Two cautions attach to the precision of this
interval. First, it is derived from a single
realisation -- the energy trajectory of one
civilisation -- and therefore quantifies the
sampling uncertainty \emph{within} our own
history, not the inaccessible dispersion of growth
rates across the population of civilisations that
the Kardashev framework ultimately concerns. Ours
is, for now, the only example on which to base a
Kardashev law. Second, even within this single
trajectory the annual observations are not
independent: the year-on-year growth rates are
positively autocorrelated ($\rho_1=0.35$;
Section~\ref{sec:independence}) and the residuals
about the fitted trend more strongly so; the
sixty data points therefore carry appreciably less
information than sixty independent, identically
distributed (i.i.d.)\ samples would. The credible
interval quoted above, which assumes independent
errors, is correspondingly optimistic, and an
autocorrelation-aware error model would widen it.
Neither caution alters the qualitative conclusions
-- the one-percent rate remains decisively
excluded -- but both temper any over-literal
reading of the $\pm0.03\%$ figure.

The Widely Applicable Information Criterion
(WAIC) \citep{Watanabe2010} comparison yields
$\Delta\mathrm{WAIC} = \mathrm{WAIC}_{\mathrm{exp}}
- \mathrm{WAIC}_{\mathrm{linear}} \approx 4$.
The model with the lower WAIC is the one
favoured, so a positive difference favours the
linear form.
The standard error of this difference, however,
is large: computed from the pointwise predictive
densities \citep{Vehtari2017},
$\mathrm{SE}(\Delta\mathrm{WAIC})\approx18$, so
the difference lies well within one standard
error of zero
($\Delta\mathrm{WAIC}/\mathrm{SE}\approx0.2$).
The slight numerical edge to the linear model is
therefore not statistically significant: on the
observed baseline, model selection prefers neither
functional form. Both carry
$K=2$ free parameters, so the Akaike and
Bayesian information criteria penalise them
identically. Together, the near-identical $R^2$
values ($\Delta R^2<0.001$) and the consistency
of $\Delta\mathrm{WAIC}$ with zero confirm the
statistical statement of Kardashev's Conundrum:
both functional forms fit the data equally well on
the observed baseline, so
the choice of which functional form to
extrapolate is underdetermined by the fit.

The complete statistical picture is:
\begin{enumerate}
\item The one-percent Kardashev exponential is
falsified by the data (poor fit; posterior
$r=2.01\%$, well outside the one-percent prior).
\item The linear model and the free-$r$
exponential have near-identical in-sample
performance ($\Delta R^2 < 0.001$), and neither is
significantly preferred by WAIC
($\Delta\mathrm{WAIC}\approx4$, with
$\mathrm{SE}(\Delta\mathrm{WAIC})\approx18$).
\item The year-over-year differences $\Delta P$
are non-Gaussian (Shapiro--Wilk $W=0.925$,
$p=0.0014$), exhibiting negative skewness driven
by identifiable historical shocks.
\end{enumerate}

The data therefore reject the fixed one-percent
rate and do not falsify exponential growth in
general. The two data-adequate descriptions,
chosen for parsimony, fail for complementary
reasons: the linear model reaches the Type~II
threshold only after a cosmologically absurd
$\sim10^5$ Hubble times (Eq.~\ref{eq:textrap}),
while the free-$r$ exponential, though it would
reach the threshold some 1.5~kyr from the
present, is excluded as a \emph{sustained}
trajectory on physical rather than statistical
grounds. Exponential energy growth is
self-terminating: at one percent per year or
faster, the accompanying waste heat renders an
Earth-like planet uninhabitable within
$\lesssim10^3$~yr \citep[see
Section~\ref{sec:technosignatures}]{BalbiLingam2025},
and, more generally, growth within any finite
resource base must saturate, the exponential
being merely the early-time limit of bounded
(logistic) dynamics. For the two forms
considered, therefore, neither yields a
physically coherent path to the energy-scaling
Type~II threshold, and other trajectories
consistent with the record could lead to very
different long-term behaviour. This is what the
Conundrum shows: the observed terrestrial energy
trajectory does not uniquely determine the
long-term path toward the Type~II threshold, and
the canonical exponential extrapolation is
physically problematic.

One conclusion, however, holds whatever the
functional form. The habitability ceiling of
\citet{BalbiLingam2025} is a threshold in power,
of order $10^{15}$~W for an Earth-like planet,
eleven to twelve orders of magnitude below
$L_\odot$; any 
trajectory whose waste heat is dissipated on the
home planet therefore crosses that ceiling before
it reaches the Type~II threshold. The growth law
sets the times of the two crossings but not their
order. A trajectory growing faster than linearly
but slower than exponentially, such as a
stretched exponential, crosses both at
intermediate times and in the same order; no
growth law, however finely tuned, postpones the
loss of habitability until after the Type~II
threshold, since that requires relocating energy
dissipation off the home planet, the transition
discussed in Section~\ref{sec:technosignatures}.

\subsection{The KSN Resolution: Renormalisation to $\mathrm{J\,Hash^{-1}}$}
\label{sec:ksn}

The energy record does not select a growth law for
$P(t)$, and no choice of growth law alters the
order in which a planet-bound trajectory crosses
the habitability ceiling and the Type~II
threshold (Section~\ref{sec:mcmc}). We therefore
do not attempt to resolve the Conundrum by
fitting a different function to $P(t)$; we turn
instead to the choice of state variable, for the
reasons that follow.

\paragraph{Dimensional incompleteness} The
Kardashev scale measures energy throughput,
$P\,[\mathrm{W}=\mathrm{J\,s^{-1}}]$, but makes no
reference to what that energy
accomplishes. As \citet{Sagan1973} noted, a
civilisation that wastes energy inefficiently
is rated equally to one that channels the same
power into information-rich computation.
The scale is therefore dimensionally incomplete as
a measure of civilisational
advancement: it conflates energy expenditure with
technological capability. That a scale based on
power alone is incomplete
has been argued independently on other grounds:
\citet{Swoboda2026} ranks multi-planetary
expansion by the useful exergy and other benefits
it produces per unit of perturbation imposed on
the host system, and \citet{JiangDas2026}
combine energy use, information processing,
construction mass and population in a weighted
composite index, bounding information processing
by the Landauer limit. Like the first, the KSN
variable is a ratio between a cost and a benefit,
here energy and computation, and like the second
it draws on the Landauer limit. It differs from
both in needing no weights, since both of its
terms are measured (the energy from published
statistics, the computation from a publicly
auditable record), and from the
second in applying the Landauer limit as a floor
under a measured quantity, whose approach to it
can be followed in time
(Section~\ref{sec:timescales}), rather than as a
ceiling on the computation a star could power.

\paragraph{The Landauer connection} The
fundamental physical link between energy and
information is the Landauer limit
\citep{Landauer1961,Bennett1982}: the minimum
energy required to erase one bit of information at
temperature $T$ is $k_B T \ln 2 \approx
2.85\times10^{-21}$~J at room temperature. The
ratio $P/H$ -- global energy production divided by
the Bitcoin \emph{proof-of-work} hashrate -- has
dimensions of energy per hash, and is an empirical
measure of how efficiently a civilisation converts
energy into irreversible computational work of the
measured class. A single hash is not a single bit
erasure. A compression of the Secure Hash
Algorithm SHA-256 comprises 64 rounds
\citep{FIPS180} of roughly $2\times10^{3}$
two-input gate operations, each of which erases
about one bit in irreversible logic, and a hash
attempt requires two compressions, so a double
SHA-256 hash erases a number
$N_b\approx2\times10^{5}$ of bits on classical
irreversible hardware. The thermodynamic floor on
$B=P/H$ is therefore not the Landauer bound itself
but a multiple of it, $B_{\min}=N_b\,k_B T\ln
2\approx6\times10^{-16}$~J per hash, some five
orders of magnitude above the single-bit bound;
equivalently, $B/N_b$, the global energy per bit
erased by the measured class, is bounded below by
$k_B T\ln 2$ for any physical substrate. The count
$N_b$, by contrast, is a property not of the hash
function alone but of the classical irreversible
logic that evaluates it: reversible logic, which
undoes its intermediate steps instead of erasing
their results, would lower it. As the energy per
hash of a civilisation's hardware approaches this
floor, the civilisation approaches the
thermodynamic limit of its computation. The
variable $B$ thus carries a physical meaning
absent from $P(t)$ alone.

\paragraph{The renormalisation} We define the
KSN state variable as
\begin{equation}
\label{eq:ksn}
B(t) = \frac{P(t)}{H(t)}\quad
[\mathrm{J\,Hash^{-1}}],
\end{equation}
where $H(t)$ is the yearly average Bitcoin
network hashrate in $\mathrm{Hash\,s^{-1}}$ and
$\mathrm{J\,Hash^{-1}}$ is the KarNak unit.
We use the Bitcoin hashrate \citep{Nakamoto2009}
because it is the dominant and longest-running
publicly auditable, unforgeable, and continuously
updated global record of any class of
computation, fulfilling the requirement of
\citet{Sagan1973} that the Kardashev state
variable be sensitive to the ``security of vast
quantities of information.'' Proof-of-work is
one specialised, economically motivated
computational workload; it is not the total
computation performed by civilisation, and $B$
is accordingly an observational proxy for one
measurable class of global computation rather
than a universal measure of civilisational
information efficiency. The \emph{choice} of this
denominator is a modelling decision rather than a
forced one -- other denominators (global
floating-point throughput, total computing
energy), were they auditable, would yield
different trajectories -- but once the
denominator is fixed the renormalisation
introduces no fitted parameter, being a ratio of
two physical quantities measured from separate
records.

\paragraph{Statistical motivation} The variable transformation $P \to B = P/H$ is not a
Box--Cox power transformation of $P$ \citep{BoxCox1964}, which would remain within the
family of functional forms fitted to the original variable. It is instead a ratio
normalisation by an external physically motivated quantity, taking the analysis outside
the Box--Cox family. The KSN variable $B(t)$ is our proposal for an information-energy state variable of the Kardashev scale.

\paragraph{Concurrent work} Complementary work
from the quantum-computing perspective has
recently examined Bitcoin mining as a
Kardashev-scale problem
\citep{DallaireDemers2026}, demonstrating that
quantum-accelerated Bitcoin mining at current
network difficulty would require a fleet of
$\sim10^{23}$ physical qubits drawing
$\sim10^{25}$~W -- resources approaching the
Kardashev Type~II threshold. Their estimate shows
that quantum mining is impractical: the saving in
the number of hash evaluations is outweighed by
the cost of computing each one on error-corrected
quantum hardware and by the number of separate
machines needed to find a block within the
ten-minute block interval. Were it achieved,
quantum mining would also break the link between
the hashrate inferred from the difficulty and the
hashes physically evaluated, since a quantum
computer needs fewer evaluations of the hash to
find a block than the difficulty implies; given
the resources it requires, however, it lies
outside the record analysed here, which counts
only hashes evaluated on classical irreversible
hardware. Our analysis is methodologically
distinct -- we use the Bitcoin hashrate as an
observational proxy for one measurable class of
global computation and renormalise the Kardashev
state variable accordingly -- but shares the use
of proof-of-work infrastructure as a natural
bridge between civilisational energy and
irreversible computation.

\paragraph{Adjacent technosphere literature} A parallel research programme has
approached the limits of civilisational growth from the perspective of technosignatures
and technosphere modelling. \citet{Wright2022} synthesise the longstanding SETI-community
argument that technosignatures may be more abundant, longer-lived, and more detectable
than biosignatures, motivating renewed interest in quantitative frameworks for
civilisational scale. \citet{HaqqMisra2025a} developed ten self-consistent scenarios for
Earth's technosphere over a thousand-year horizon -- three with zero-growth stability,
two that have collapsed, one that oscillates, and four that continue to grow --
illustrating that conventional Kardashev-style continuous-growth extrapolation captures
only a subset of plausible long-term futures. \citet{HaqqMisra2025b} argue that the
Kardashev scale is better understood as a ``luminosity limit'' -- the maximum capacity
for a civilisation to harvest luminous stellar energy within a given spatial domain --
and that thermodynamic efficiency keeps any luminosity-limited technosphere strictly
below this theoretical maximum. \citet{BalbiLingam2025} analyse, via simple thermodynamic
models, the constraints that unavoidable waste-heat generation imposes on the long-term
habitability of Earth-like planets hosting technospheres in persistent exponential
growth, demonstrating that the loss of habitable surface conditions can be expected on
timescales of $\lesssim10^3$~yr from the onset of the exponential phase, for an assumed
annual energy-consumption growth rate of $\sim1\%$ -- the same conservative Kardashev
rate that the present work statistically falsifies. \citet{Cirkovic2018} situates such
constraints within a four-category taxonomy of Fermi-paradox solutions, whose
``logistic'' category groups hypotheses in which Galactic sophonts exist and are in
principle detectable but the cost of maintaining detectability over astrophysical
timescales is prohibitive. We draw these threads together in
Section~\ref{sec:technosignatures}, which states
the assumptions under which the results of
Sections~\ref{sec:standard}--\ref{sec:mcmc}
supply a specific mechanism within that
logistic category.

\begin{figure}[htbp]
\centering
\includegraphics[width=0.8\linewidth]{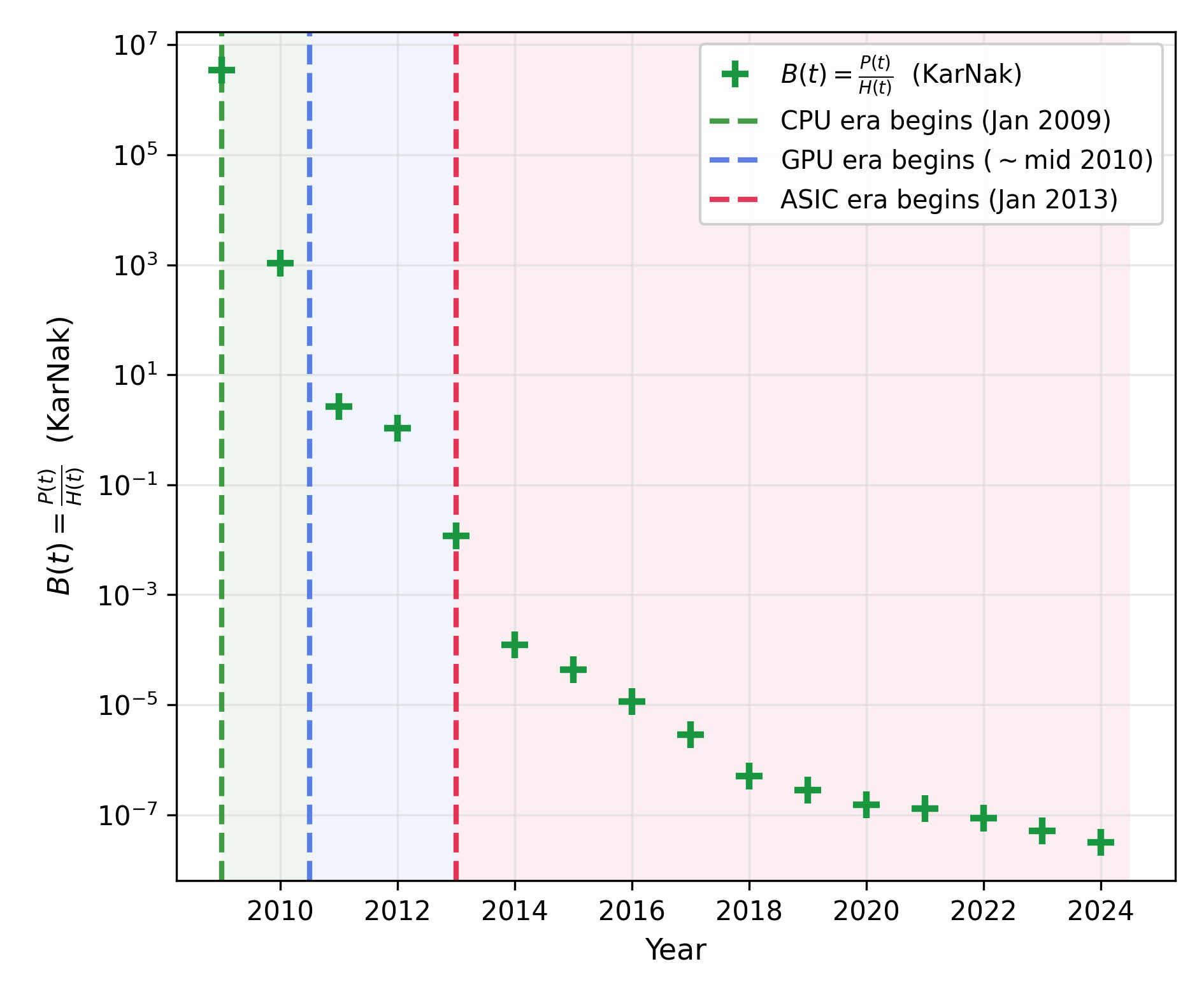}
\caption{The KSN state variable $B(t)=P(t)/H(t)$ in KarNak units ($\mathrm{J\,Hash^{-1}}$)
for 2009--2024, spanning 14 orders of magnitude. The steeply decreasing trend combines the
improving energy efficiency of proof-of-work
hardware, from the CPU era (2009) through the
GPU era (2010--2012) to the ASIC era
(2013--2024), with the growing share of global
energy devoted to the Bitcoin network. Structural
breaks between eras preclude a single-exponential description; a formal fit is deferred
to a future paper. Each annual hashrate is computed from the blocks
mined in that year (Section~\ref{sec:data}); its
uncertainty, of order one percent, is far
smaller than the plotting symbols on this
fourteen-decade logarithmic ordinate, so error
bars are not shown.}
\label{fig:ksn}
\end{figure}

Figure~\ref{fig:ksn} shows $B(t)$ for 2009--2024,
spanning 16 annual data points and 14 orders of
magnitude in $B$. Because $P(t)$ varies by only
$\sim30\%$ over this window while $H(t)$ grows by
fourteen orders of magnitude, $B=P/H$ is
numerically dominated by the hashrate denominator.
The decline of $B(t)$ combines the improving
energy efficiency of the hardware with the growing
share of global energy devoted to the network. The
data display a steeply decreasing trend from the
CPU era (2009) through the graphics processing
unit (GPU) era (2010--2012) to the ASIC era
(2013--2024). The structural breaks between these
eras preclude a single-exponential description of
$B(t)$; a formal era-segmented functional fit is
deferred to a future paper.

\subsection{Civilisational Timescales on the KSN Scale}
\label{sec:timescales}

Table~\ref{tab:timescales} summarises the
Type~II timescales implied by each model. The
Kardashev and Sagan exponential models give
timescales of order $10^3$~yr. A crossing is
built into their functional form: any exponential
$P=P_0 e^{rt}$ with $r>0$ reaches any finite
threshold at some finite time. The Type~II
crossing year is therefore set by the assumed
growth rate $r$ and carries no information beyond
that rate; the testable quantity is the rate
itself (Section~\ref{sec:falsifiability}). The
linear model, as shown in
Eq.~(\ref{eq:textrap}), gives a timescale that is
cosmologically absurd. On the KSN scale, the
concept of a Type~II threshold must be
re-expressed in terms of $B=P/H$: a Type~II KSN
civilisation is one
whose hardware energy per hash approaches its
thermodynamic floor, the per-hash Landauer limit
$N_b\,k_B T\ln 2$ (a hardware-dependent multiple
of the single-bit bound). This reframing is
consistent with the original intent of
\citet{Kardashev1964} and the
information-theoretic extension proposed by
\citet{Sagan1973}.

\paragraph{Timescale for approach to the floor}
A direct extrapolation of the trend in
Figure~\ref{fig:ksn} would bring $B$ to
$B_{\min}$ between the 2040s and the 2060s,
depending on whether the trend is taken over
the whole ASIC era or over its last five years,
but the trend cannot physically continue that
far, with today's hardware or with any later
hardware built on fast, reliable irreversible
logic. The energy
per hash of the hardware in use,
$\varepsilon_{\mathrm{hw}}$, is by definition
the power drawn by the network divided by the
hashrate. With $f$ the share of global energy
devoted to the measured class, the network
draws the power $fP$, so that
$H=fP/\varepsilon_{\mathrm{hw}}$ and
$B=\varepsilon_{\mathrm{hw}}/f$ identically.
Reaching $B_{\min}$ would require
$f=\varepsilon_{\mathrm{hw}}/B_{\min}$, which is
$10^4$ to $10^5$ at present hardware efficiency
and would still be nearly two orders of
magnitude above one at the practical bound of
such logic (given below), but a
share cannot exceed one: $B$ cannot fall below
$\varepsilon_{\mathrm{hw}}$, however large the
world's primary power. The identity instead
splits the
seven to eight orders of magnitude between
$B(2024)$ and $B_{\min}$ into this hardware
factor, $\varepsilon_{\mathrm{hw}}/B_{\min}$, and
an allocation factor, $1/f\approx10^3$: the
network's electricity use, about 138~TWh per
year \citep{Neumueller2025}, is some
$8\times10^{-4}$ of world primary energy, so
that $\varepsilon_{\mathrm{hw}}\approx
2.5\times10^{-11}$~J per hash in 2024.

The hardware factor has a physical trend. The
energy per hash of the best available hardware
fell from about $10^{-8}$~J for the first ASICs
of 2013 to about $7\times10^{-11}$~J by 2017,
once they had reached the most advanced process
node \citep{Taylor2017}, and to
$1.35\times10^{-11}$~J by 2024
\citep{Bitmain2024}; since reaching that node it
has improved by about one order of magnitude per
decade, close to the post-2000 rate for
computing hardware generally
\citep{KoomeyNaffziger2015}. For fast, reliable
irreversible logic the practical bound on
$\varepsilon_{\mathrm{hw}}$ lies well above
$k_B T\ln 2$ per operation, at the signal energy
that reliable switching requires against thermal
noise, some tens of $k_B T$ per operation
\citep{Agarwal2016}, or a few times
$10^{-14}$~J per hash. At the present rate,
applied to the hardware in use, that
level is reached in about three decades, over
which, at fixed allocation, $B$ falls by a
further two and a half to three orders of
magnitude, to a few times $10^{-11}$~J per hash.
Within the bound $f\le1$, the allocation factor
is set by economics, by what the measured
class can spend on energy, and the model makes
no prediction for it; once the hardware factor
saturates, $B$ scales as $1/f$. $B_{\min}$ is
the floor of irreversible logic, reached only in
the limit of slow operation \citep{Berut2012};
hardware that must switch quickly and reliably
stops nearly two orders of magnitude above it,
and only reversible logic \citep{Bennett1982},
which erases no intermediate bits and so is not
bound by it, can go further. On irreversible
hardware the measured $B$
could approach $B_{\min}$ only if, in addition,
the allocation approached unity. The hardware
factor thus lies within about three decades of
its practical floor for irreversible logic,
against the $10^5$ Hubble times of the
energy-scaling threshold on the linear
description; the remainder requires slow
operation or reversible
logic, for which there is no trend to
extrapolate, and the allocation factor is set
by economics.

\begin{table}[htbp]
\centering
\caption{Type~II civilisation timescales under
each model. All timescales are measured from
1964. $H_0^{-1}\approx14$~Gyr is the Hubble time. The
MCMC exponential uses the posterior mean
$r=2.01\%~\mathrm{yr^{-1}}$. For the KSN model
the condition is placed on the hardware energy per
hash, $\varepsilon_{\mathrm{hw}}$, whose Landauer
floor is $N_b\,k_B T\ln 2\approx6\times10^{-16}$~J
per hash; in place of a year, the table gives the
decade in which, at the present rate of
improvement, $\varepsilon_{\mathrm{hw}}$ reaches
the practical floor of irreversible logic, nearly
two orders of magnitude above that value.}
\label{tab:timescales}
\small\setlength{\tabcolsep}{4pt}
\begin{tabular}{@{}llll@{}}
\hline
Model & Growth rate & Type~II year (CE) &
$\Delta t/H_0^{-1}$ \\
\hline
Kardashev 1964 (fixed $r$) &
$1.00\%~\mathrm{yr^{-1}}$ & $\sim 5200$ &
$\sim 2\times10^{-7}$ \\
Exp.\ free $r$ (MCMC mean) &
$2.01\%~\mathrm{yr^{-1}}$ & $\sim 3500$ &
$\sim 1\times10^{-7}$ \\
Linear OLS &
$2.44\times10^{11}~\mathrm{W\,yr^{-1}}$ &
$\sim 1.6\times10^{15}$ & $\sim 1.1\times10^{5}$ \\
KSN ($B=P/H$) &
$\varepsilon_{\mathrm{hw}}\to N_b\,k_B T\ln 2$ &
2050s & $\sim 6\times10^{-9}$ \\
\hline
\end{tabular}
\end{table}

\begin{table}[htbp]
\centering
\caption{Comparison of energy-production
baselines and Type~II civilisation timescale
predictions. $P_0$ is the average continuous power
at the start of each analysis, converted from the
reported annual energy via Eq.~(\ref{eq:conv}).
$H_0^{-1}\approx14$~Gyr is the Hubble time.}
\label{tab:baselines}
\small\setlength{\tabcolsep}{4pt}
\begin{tabular}{@{}lll@{}}
\hline
Source & $P_0$ & Type~II timescale \\
\hline
\citet{Kardashev1964} (1964 estimate) &
$4$~TW &
$\sim 3.2$~kyr at $r=1\%~\mathrm{yr^{-1}}$ \\
OWID \citep{Ritchie2022} (1965, this work) &
$4.95$~TW &
$\sim 1.1\times10^{5}\,H_0^{-1}$ (linear OLS) \\
\hline
\end{tabular}
\end{table}

\subsection{The Independence Property as a Further Objection to the Exponential}
\label{sec:independence}

A complementary objection arises from the theory of stochastic processes. The deterministic
series $P(t)=P_0(1+x)^t$ is the mean path of a multiplicative process $P_t=G_t\,P_{t-1}$
whose annual growth factors $G_t$ have mean $1+x$. The model's essential structural
assumption is that the $\{G_t\}$ are independent and identically distributed; equivalently,
the log-increments $\xi_t\equiv\ln(P_t/P_{t-1})$ are i.i.d., so that $\ln P_t$ is a random
walk with constant drift. Such a process is \emph{memoryless} in the precise sense that,
conditional on the present, the next increment is independent of the entire history that
produced it:
\begin{equation}
\label{eq:memoryless}
\Pr\!\left(\xi_{t+1}\le z \,\middle|\, \xi_t,\xi_{t-1},\dots,\xi_1\right)
   = \Pr\!\left(\xi_{t+1}\le z\right).
\end{equation}
The model thus embeds no channel through which the trajectory's past can shape its future
beyond fixing the current level.\footnote{This process-level memorylessness is
the stochastic-process counterpart of the
memoryless property $P(X>a+b\mid X>b)=P(X>a)$,
which \emph{uniquely} characterises the geometric
distribution among discrete laws and the
exponential among continuous ones; in both
settings the defining feature is that the past
conveys no information about the future.
Memorylessness is therefore a highly exceptional
property.}

The record departs from this model in three
ways. First, the log-increments are not
independent draws from a single distribution:
the lag-one autocorrelation of $\xi_t$ is
$\rho_1=0.35$, and a Ljung--Box test rejects the
hypothesis of no autocorrelation at every lag
from one to four ($Q=7.8$--$12.8$, $p<0.02$;
Durbin--Watson $=1.25$; runs-test $z=-2.5$,
$p=0.014$). These tests do not show whether the
departure comes from the persistence of shocks
from one year to the next or from a change in
the mean growth rate over the record; either
violates the model's assumption. Second, the
year-over-year differences $\Delta P$ are
significantly non-Gaussian and negatively skewed
(Shapiro--Wilk $W=0.925$, $p=0.0014$; skewness
$-0.664$; Figure~\ref{fig:deltaP}), with the
heaviest negative excursions traceable to
identifiable exogenous shocks (2009, 2020) rather
than to symmetric random draws; this shows that
the shocks are not Gaussian, though not that
they depend on one another. Third, and most
simply, the series contains years of outright
negative growth (e.g.\ 2009, 2020), which the
strictly increasing model $(1+x)^t$ with $x>0$
cannot represent at any parameter value. The
independence property is therefore not an
abstract technicality but a testable claim about
the generative process, and the evidence is
against it: the log-increments of the record are
not the independent, identically distributed
draws that the constant-rate model requires.

Before discussing falsifiability and limitations, we note the most immediate consequence
of the foregoing analysis. Whatever the growth law, a planet-bound
trajectory crosses the waste-heat ceiling of
habitability before it reaches the energy-scaling
Type~II threshold (Section~\ref{sec:mcmc}).
Type~II energy-harvesting technosignatures such
as Dyson spheres are therefore unlikely to arise
from a scaled-up terrestrial trajectory, a
candidate physical explanation for their
continued non-detection that we treat in full in
Section~\ref{sec:technosignatures}.

% =====================================================================
\section{Discussion}
\label{sec:discussion}

\subsection{A Physical Explanation for the Non-Observation of Type~II Technosignatures}
\label{sec:technosignatures}

The results of Section~\ref{sec:analysis} carry a direct and, to our
knowledge, previously unquantified consequence for SETI. We state it at the head of the
discussion because it is the result most likely to be of interest to the observational
technosignature community, and because the assumptions it requires can be stated
explicitly, which we do first.

The inference rests on three assumptions, which we
state before drawing it.
(i)~\emph{Representativeness}: the terrestrial
energy record is taken as representative of the
planet-bound phase of a technological
civilisation's growth. This cannot be tested with
a single example, and an extraterrestrial
civilisation need not follow the terrestrial
trajectory before constructing a Dyson swarm.
(ii)~\emph{Planet-bound dissipation}: growth is
assumed to remain confined to the home planet, so
that the waste heat of energy production is
dissipated there. A civilisation that expands into
space, or that exploits stellar energy
incrementally from the outset, relaxes this
assumption and with it the habitability ceiling
that bounds the planet-bound phase.
(iii)~\emph{Energy-scaling path}: the route to
Type~II is assumed to run through the scaling of
general energy consumption, as in the terrestrial
record, rather than through a demand for
computation that captures stellar energy directly.
A civilisation that uses stellar energy primarily
for computation relaxes this assumption, and its
trajectory is described by the informational
variable of Section~\ref{sec:ksn} rather than by
$P(t)$ alone. Under (i)~to~(iii), the record
establishes the following. On either of the two
growth forms analysed, the energy-scaling
threshold is not reached before the habitability
ceiling is crossed, and for any planet-bound
trajectory that ordering holds whatever the growth
law, because the habitability ceiling of
\citet{BalbiLingam2025}, a total waste-heat power
of order $10^{15}$~W for an Earth-like planet
(their $K_{\max}\approx0.85$-$0.9$ on Sagan's
interpolated scale \citep{Sagan1973}), lies eleven
to twelve orders of magnitude below $L_\odot$
(Section~\ref{sec:mcmc}). The planet-bound phase
therefore ends within one to two orders of
magnitude of present power, whatever its form. A
civilisation that reaches the energy-scaling
Type~II threshold must have relocated its energy
dissipation off its home planet, and Dyson-type
structures, if they exist, are the product of that
transition rather than of a scaled-up terrestrial
trajectory. The argument that follows bears on how
likely that transition is; it does not exclude it.

The one-percent geometric model is rejected
outright (Section~\ref{sec:standard}); the data
are described equally well by a linear model and
a free-rate exponential at
$r\approx2\%~\mathrm{yr^{-1}}$, neither
significantly preferred by WAIC
($\Delta\mathrm{WAIC}\approx4$, well within its
standard error; Section~\ref{sec:mcmc}). Under
the two other energy-accounting conventions the
difference stays below two standard errors; it
is largest on the total energy supply, where the
linear form is favoured by 1.8 standard errors
(Section~\ref{sec:limitations}). On the
linear description, the energy-scaling Type~II
threshold $P\to L_\odot$ is reached only after
$t^\ast\approx1.6\times10^{15}~\mathrm{yr}
\approx1.1\times10^{5}\,H_0^{-1}$
(Eq.~\ref{eq:textrap}) -- a timescale that
exceeds the main-sequence lifetime of the host
star by some five orders of magnitude. The
implication is not merely that Type~II is
distant: it is that the star a Dyson sphere would
enclose evolves off the main sequence, exhausts
its hydrogen, and is gone long before a
civilisation growing its energy budget at the
observed rate, on its home planet, could
approach that star's luminosity. For such a
trajectory the energy-scaling Type~II threshold
is not a destination at all; on the exponential
description, the waste-heat ceiling of
habitability is crossed long before the
threshold is reached (Section~\ref{sec:mcmc}).
A Type~II energy-harvesting technosignature (a
Dyson sphere, or any equivalent full-luminosity
collector) is therefore unlikely to arise from a
scaled-up terrestrial trajectory; how common
such structures are in the Galaxy depends on how
often civilisations make the off-planet
transition of assumption (ii), which the
terrestrial record does not constrain.

This furnishes a candidate physical explanation
for the outcome of the observational Dyson-sphere
search programme. The IRAS whole-sky upper limit
of \citet{Carrigan2009}, the $\hat{G}$ infrared
search of \citet{Wright2014}, and the
Gaia--2MASS--WISE candidate search of
\citet{Suazo2024} have each returned no confirmed
Type~II technosignature. Such null results are
conventionally attributed to the rarity of
civilisations, the transience of the Type~II
phase, or the sensitivity limits of the surveys.
Our analysis adds a more elementary possibility:
that the search target itself, a structure defined
by full-stellar-energy scaling, corresponds to a
civilisational state that no planet-bound
technosphere attains, and that is reached only by
relocating energy dissipation off the home world.

This conclusion was anticipated, on qualitative
grounds, by \citet{Ivanov2020}, who argued that
present-day searches are tuned to find only
civilisations that share our own
``more-is-better'' growth philosophy, and that the
failure of Dyson-sphere searches may indicate the
inadequacy of that philosophy rather than an empty
Galaxy. The present work supplies the missing
quantitative step: a direct statistical
falsification, against the observed energy record,
of the fixed one-percent rate that Kardashev
extrapolated to Type~II, together with the
physical argument of Section~\ref{sec:mcmc}, valid
whatever the growth law, that planet-bound growth
reaches the waste-heat ceiling of habitability
before the energy-scaling Type~II target.

Within the four-category taxonomy of Fermi-paradox
solutions of \citet{Cirkovic2018}, this result
belongs to the ``logistic'' category --
civilisations exist and are in principle
detectable, but the cost of attaining or
sustaining a given detectable state over
astrophysical timescales is prohibitive. What is
striking is that three logically independent
arguments now converge on the same conclusion for
energy-scaling Type~II states. (i)~The present
statistical analysis shows that the canonical
one-percent growth law is not supported by the
terrestrial energy record, and that the observed
terrestrial trajectory does not uniquely determine
the long-term path toward the Type~II threshold.
(ii)~The waste-heat habitability constraint of
\citet{BalbiLingam2025} shows that sustained
exponential growth at the very $\sim1\%$ rate
Kardashev assumed would render an Earth-like
planet uninhabitable within $\lesssim10^3$~yr;
because the constraint is a threshold in power far
below $L_\odot$, any planet-bound trajectory
crosses it before the Type~II threshold, whatever
its growth law (Section~\ref{sec:mcmc}). (iii)~The
luminosity-limit argument of
\citet{HaqqMisra2025b} shows that thermodynamic
efficiency holds any luminosity-limited
technosphere strictly below the
full-stellar-luminosity ceiling. The first
argument is statistical, the second
ecological-thermodynamic, the third
efficiency-thermodynamic; their mutual
independence is precisely what makes their
agreement significant. The canonical
energy-scaling path to Type~II is disfavoured
whether one reasons from the energy record, which
does not support its growth law, from planetary
habitability, or from the thermodynamics of energy
capture.

Two caveats bound this inference, and we state them plainly. First, it rests on
assumptions (i)~to~(iii) above and on a single civilisation observed over a single
six-decade window; growth regimes that would restore exponential scaling, a future
technological transition or expansion beyond the home planet onto a stellar or
interstellar resource base, are not excluded, and \citet{HaqqMisra2025a} explicitly
construct self-consistent thousand-year futures in which growth continues. Second, the claim is specific to energy-scaling technosignatures:
it places no constraint on technosignatures defined along other axes, such as directed
radio or optical emission, atmospheric industrial products, or the information-theoretic
signatures implied by the KSN variable. What the convergence of three independent
arguments does establish is a shift in the burden of proof: a Galaxy-common population of
persistent Dyson spheres can no longer be treated as the natural expectation of the
Kardashev framework, but must instead be justified against three distinct objections, one statistical and
two physical.

The argument requires a further distinction. The Kardashev Type~II classification and
the Dyson sphere are not the same thing. Type~II is a statement about power, a rate of
energy use comparable to the luminosity of the host star, and it says nothing about how
that energy is collected; a Dyson structure is one realisation of stellar-scale energy
use, sufficient but not necessary, and Dyson's own proposal was a swarm of independent
collectors rather than a shell \citep{Dyson1960}. Between the present and the
full-luminosity limit lies a continuum of partial harvesting: swarms intercepting a small
fraction of the stellar output, or distributed orbital collectors serving computation
rather than general consumption, each corresponding to a state between Type~I and
Type~II \citep{Wright2020}. The surveys cited above are sensitive to the reprocessed
waste heat of any such structure in a way that depends on the fraction of luminosity it
intercepts, so the argument of this section bears on the full-luminosity end of that
continuum, the energy-scaling Type~II threshold, and not on partial harvesting at levels
below present survey sensitivity. A civilisation whose development is primarily
informational rather than energetic need not move along that continuum at all
\citep{Sandberg2016}. This is the distinction on which the KSN proposal rests: a
Type~II-like state can be defined without requiring a Dyson structure.

Finally, the waste-heat ceiling on planet-bound
growth does not abolish the Type~II concept; it
relocates it. Under the KSN renormalisation of
Section~\ref{sec:ksn}, the Type~II milestone is
re-expressed as the approach of the hardware
energy per hash of the measured class to its
Landauer floor, $\varepsilon_{\mathrm{hw}}\to
N_b\,k_B T\ln 2$ (Section~\ref{sec:timescales});
the measured $B=\varepsilon_{\mathrm{hw}}/f$ then
lies a factor $1/f$ above that floor, set by the
energy share of the class. A civilisation may thus
be ``Type~II'' in the information--energy sense --
performing irreversible computation at the
thermodynamic floor -- without ever enclosing a
star. The non-observation of energy-scaling
Type~II technosignatures and the survival of the
Type~II concept are therefore consistent: what
fails is one dimensionally incomplete realisation
of the milestone, while the scale's classificatory
structure is preserved on a state variable that
our own civilisation is measurably traversing
(Figure~\ref{fig:ksn}). This is the sense in which
we call the KSN model a resolution of the
Conundrum rather than a rescaling: the Type~II
milestone survives on a measured variable, on
which it is approached within decades
(Section~\ref{sec:timescales}) rather than after
the $10^5$ Hubble times of the linear description.

\subsection{Falsifiability of the KSN Model}
\label{sec:falsifiability}

A scientific model is meaningful only insofar as it generates predictions that can in
principle be refuted by data \citep{Blumer1987}. We identify four concrete falsification
criteria, the first two for the analysis of the
energy record and the last two for the KSN
variable.

First, the two descriptions of the energy
record, neither yet significantly preferred by
model selection, differ in what they hold
constant. The free-rate
exponential holds the growth rate constant, and
its posterior value,
$r=2.01\%\pm0.03\%~\mathrm{yr^{-1}}$, is a
quantitative prediction: if sustained
observation of global energy production over the
coming decades yields a growth rate consistently
outside the 95\% credible interval
$[1.94\%, 2.08\%]$, the exponential description
is falsified. The linear description holds the
annual increment constant, at
$b=2.44\times10^{11}~\mathrm{W\,yr^{-1}}$, so
that the growth rate falls as $P$ grows; a
sustained change in the annual increment would
falsify it. Both tests require energy measured
under the same accounting convention as the series
used here, because the fitted values depend on the
convention (Section~\ref{sec:limitations}).
Neither outcome would remove the motivation for a
measure other than power, since a planet-bound
trajectory crosses the waste-heat ceiling of
habitability before it reaches the Type~II
threshold whatever its growth law
(Section~\ref{sec:mcmc}).

Second, the year-over-year energy increments
$\Delta P$ are non-Gaussian (Shapiro--Wilk
$W=0.925$, $p=0.0014$), with their largest
negative excursions driven by identifiable
historical events. This is a testable claim: if
future decades produce a $\Delta P$ distribution
consistent with Gaussian increments, it is
falsified.

Third, the decline of $B$ should end. By the
estimate of Section~\ref{sec:timescales}, the
hardware factor $\varepsilon_{\mathrm{hw}}$
reaches the practical floor of irreversible
logic, a few times $10^{-14}$~J per hash, in the
2050s at the present rate of improvement;
$B=\varepsilon_{\mathrm{hw}}/f$ then scales as
$1/f$ and, if the allocation stays near its
present value, levels off at a few times
$10^{-11}$~J per hash. A sustained rise in the
energy per hash of the best available hardware
would falsify the trend on which this estimate
rests, and fast, reliable irreversible hardware
operating below the practical floor would show
that $N_b$ or the signal energy required for
reliable switching has been overestimated. A
hardware factor that levels off above the
practical floor would not contradict the
framework; it would locate the limit of the
present technology.

Fourth, since $f\le1$, the measured $B$ cannot
fall below the hardware factor, which on
classical irreversible hardware cannot fall below
$B_{\min}$. A measured $B$ below $B_{\min}$
would therefore show either that each hash erases
fewer bits than $N_b$, as it would with
reversible logic, or that $H$ counts hashes that
were never evaluated, as it would under quantum
mining (Section~\ref{sec:ksn}). The second would
appear first as a rise in $H$ that neither the
energy use of the network nor the efficiency of
its hardware can account for; in either case, $B$
could not be compared across the change without
re-evaluating $N_b$.

\subsection{Limitations}
\label{sec:limitations}

Several limitations of the present analysis deserve explicit acknowledgement.

A methodological choice concerns the
energy-accounting convention. Energy statistics
differ in how they count the electricity generated
from nuclear, hydro, wind and solar sources. The
OWID series used here follows the substitution
convention
(Section~\ref{sec:data}), which counts it as the
fossil fuel that would have been needed to
generate it, some 2.4 to 2.8 times the electricity
itself. The direct convention counts the
electricity as generated. A third convention
counts nuclear electricity as the heat released
in the reactor, and hydro, wind and solar
electricity as generated; it underlies the total
energy supply that the EI and OWID now report
\citep{Ritchie2026}. For the electricity from
these four sources, the direct and the third
conventions count only energy that was actually
produced, as electricity or as reactor heat, and
no fossil-fuel equivalent.

The analysis of Section~\ref{sec:analysis} uses
the series as published in the release of
27 April 2026, which predates OWID's change to
total energy supply, and we repeated the linear
and exponential fits and their comparison under
the direct and the third conventions. Recounted
under the direct convention, the world total is
4 percent lower in 1965 and 10 percent lower in
2024 than in the series used here; least-squares
fits give
$b=2.18\times10^{11}~\mathrm{W\,yr^{-1}}$ for the
linear model, reaching the Type~II threshold after
$\sim1.3\times10^{5}\,H_0^{-1}$, and
$r=1.94\%~\mathrm{yr^{-1}}$ for the exponential.
In the total energy supply of the OWID release of
10 September 2026, which follows the third
convention, the world total is 4 percent lower
in 1965 and 7 percent lower in 2024 than in the
series used here, and
the corresponding values are
$b=2.30\times10^{11}~\mathrm{W\,yr^{-1}}$,
$\sim1.2\times10^{5}\,H_0^{-1}$ and
$r=1.96\%~\mathrm{yr^{-1}}$. Under each of the
three conventions the one-percent rate is
excluded, model selection shows no significant
preference between the linear and the exponential
form, and neither form yields a physically
coherent path to the Type~II threshold, so the
Conundrum holds under all three. How strongly one
form is favoured over the other does depend on the
convention: $\Delta\mathrm{WAIC}=4\pm18$ under the
substitution convention and $-4\pm16$ under the
direct convention (positive values favour the
linear form), against $29\pm16$ on the total
energy supply ($27\pm17$ when the release of
27 April 2026 is recounted under the third
convention). On the total energy supply the linear
form is thus favoured by 1.8 standard errors,
below the two standard errors that we take as a
significant preference. The choice of convention
leaves untouched the dimensional-incompleteness
argument that motivates the KSN renormalisation.

A further methodological consideration concerns
per-year measurement uncertainties. The OWID
dataset reports point estimates only; neither
the Energy Institute nor the EIA publishes
formal per-year uncertainty bands on global
primary-energy production. Methodological
considerations suggest that pre-1980 data carry
larger fractional uncertainty ($\sim5$--$10\%$)
than post-2000 data ($\sim1$--$2\%$) due to
incomplete reporting coverage in earlier
decades, retroactive harmonisation of accounting
methods, and imputation for missing country-year
cells. To test the sensitivity of our
conclusions to this difference in uncertainty,
we assign to each yearly value a $1\sigma$
uncertainty of 10 percent of the value in 1965,
falling linearly to 1.5 percent in 2000 and
constant thereafter (the error bars of
Figures~\ref{fig:energy_loglog} and
\ref{fig:energy_linear}), and repeat the model
fits under $1/\sigma_i^2$ weighting. The
weighted re-analysis leaves every qualitative
conclusion unchanged: the Kardashev one-percent
exponential is rejected decisively (by a
residual-sum-of-squares excess exceeding two
orders of magnitude) under any reasonable
weighting; the linear-versus-exponential WAIC
comparison remains inconclusive, although its
sign reverses
($\Delta\mathrm{WAIC}=-50\pm38$, against
$4\pm18$ unweighted); the best-fit linear slope
changes by $\approx5\%$ (to
$\approx2.56\times10^{11}$~W~yr$^{-1}$) under
the adopted weighting, as the higher-uncertainty
early points are down-weighted -- immaterial to
the rejection of the one-percent model or to the
order-of-magnitude Type~II timescale; and the
Shapiro--Wilk non-Gaussianity is driven
primarily by the 2009 and 2020 outliers in
well-measured recent decades, where any
reasonable uncertainty estimate is small. A
formal weighted MCMC with fully propagated
priors on these uncertainties is left to future
work; the present sensitivity analysis suffices
to establish that the qualitative conclusions
are robust to the adopted error model.

The Bitcoin network hashrate is the dominant and
longest-running publicly auditable, unforgeable,
and continuously updated global record of any
class of computation. We do not consider Bitcoin
as one network among many to be aggregated: hashes
of different proof-of-work algorithms are different
units of work and cannot be summed, and Bitcoin has
accounted for the largest share of the energy
expended on proof-of-work continuously since 2009.
We therefore treat Bitcoin as a representative
measurement of global proof-of-work computation
rather than as one constituent among many. What the proxy measures is made precise by the
identity $B=\varepsilon_{\mathrm{hw}}/f$ of
Section~\ref{sec:timescales}. The hardware factor
$\varepsilon_{\mathrm{hw}}$, the energy per hash
of the hardware in use, is representative of the
civilisation's computing hardware in its rate of
improvement, though not in its absolute value,
since proof-of-work hardware is fabricated at the
leading process nodes \citep{Taylor2017} and,
once there, its efficiency has improved at about
the rate of computing hardware generally
\citep{KoomeyNaffziger2015,Koomey2011}; the
allocation factor $f$, the share of global energy
devoted to the measured class, is specific to
that class and set by its economics. The proxy
therefore tracks the first factor for the
civilisation and the second for one workload
only, and it does not measure the total
computation of civilisation.

The hashrate data span only 2009--2024 -- sixteen annual data points covering
approximately one human generation. The KSN variable exhibits structural breaks between
the CPU/GPU era (2009--2012) and the ASIC era (2013--2024), reflecting qualitative changes
in the technology of proof-of-work computation rather than smooth continuous evolution. A
formal era-segmented fit to $H(t)$ is deferred to a future paper.

A further limitation concerns the production--consumption distinction. The OWID dataset
reports total primary energy production, whereas \citet{Kardashev1964} defined his scale
in terms of energy consumption. We have argued that production constitutes a valid
upper-limit proxy for consumption at the civilisational scale, and that the two track each
other closely over decadal baselines. However, the ratio of production to consumption is
not precisely unity and varies with time, introducing a systematic uncertainty that is
difficult to quantify without a separate long-baseline consumption dataset. We note that
our pull request to the OWID energy repository \citep{Ritchie2022} identified and
documented this label inconsistency between the column name
\texttt{primary\_energy\_consumption} and the actual measured quantity, which is
production.

\subsection{The Novel Observation on Kardashev's Growth Rates}
\label{sec:growthrates}

A result that has received little attention in the
literature emerges directly from re-reading
\citet{Kardashev1964}. Kardashev references two
distinct growth rates in the same paper: a
projected empirical rate of
$3$--$4\%~\mathrm{yr^{-1}}$ attributed to
\citet{Putnam1948} and projected over the
following sixty years, and a conservative
modelling assumption of $x=1\%~\mathrm{yr^{-1}}$
adopted for the Type~II timescale calculation. The
standard Kardashev model widely cited in the
literature \citep{Gray2020} uses the conservative
1\% figure, not the Putnam-derived 3--4\%
projection.

Our MCMC posterior growth rate of
$r=2.01\%\pm0.03\%~\mathrm{yr^{-1}}$ falsifies
both of Kardashev's reference values: the 1\%
modelling assumption is too low, and the Putnam
3--4\% projection is too high. The fitted rate
lies between the two values Kardashev
referenced, so the canonical Type~II timescale of
about $3{,}200$~yr, which rests on the lower
value, loses its empirical basis. The record
does not, however, falsify exponential growth as
a description of its trend: a fixed rate near
2\% describes it as well as a linear form does
(Section~\ref{sec:mcmc}), although its
year-to-year growth rates are not the
independent, identically distributed draws that
a constant-rate process would produce
(Section~\ref{sec:independence}).

\subsection{Connection to the Drake Equation}
\label{sec:drake}

The Drake equation for the number of communicating civilisations in the Galaxy includes a
longevity factor $L$ -- the average duration for which a civilisation remains detectable.
Monte Carlo approaches to estimating the probability of causal contact between
communicating civilisations \citep{Lares2020} provide a complementary framework in which
the KSN variable may eventually constrain the longevity factor. This factor carries the
largest relative uncertainty of any term in the Drake equation and is sensitive to whether
civilisations tend to self-destruct or achieve long-term stability \citep{Gray2020}.

The KSN model is relevant to the longevity factor
in the following sense: a civilisation that
approaches the Landauer limit -- that is, one
whose hardware energy per hash approaches
$N_b\,k_B T\ln 2$ -- has achieved near-optimal use
of the energy it spends on computation. A
planet-bound civilisation must hold its power
below the waste-heat ceiling of
Section~\ref{sec:technosignatures} to remain
habitable, so once it nears that ceiling its
computation can grow further only through a
falling energy per operation, a route bounded by
the Landauer floor. The KSN Type~II threshold,
defined as the hardware energy per hash
approaching its Landauer floor, is therefore not
merely an energy milestone but an
information-thermodynamic one. Whether $L$ is
large or small may depend on whether civilisations
reach this threshold before exhausting or
destabilising their energy resources -- a question
the KSN model frames quantitatively.

The longevity factor is not an abstraction: the Bulletin of the Atomic Scientists set the
Doomsday Clock at 85 seconds to midnight in 2026, the closest in its history. Whether the
approach of computing hardware to its thermodynamic floor (Section~\ref{sec:timescales})
will be accompanied by the institutional stability that a long-lived technosphere requires,
what \citet{Sagan1985} in his 1985 congressional testimony called ``a degree of
international amity which certainly doesn't exist today,'' lies beyond the scope of this
model, which supplies only the physical part of that accounting.

% =====================================================================
\section{Conclusions}
\label{sec:conclusions}

Empirically, the specific one-percent Kardashev
extrapolation is not supported by the terrestrial
energy record: tested against six decades of
global energy-production data (1965--2024), the
standard one-percent exponential model is
statistically untenable. The posterior growth rate
from MCMC inference is
$r=2.01\%\pm0.03\%~\mathrm{yr^{-1}}$ (95\%
credible interval $[1.94\%, 2.08\%]$). The
year-over-year energy differences are
significantly non-Gaussian (Shapiro--Wilk
$W=0.925$, $p=0.0014$), with negative skewness
attributable to identifiable crisis events (2009,
2020), and the annual growth rates are not the
independent, identically distributed draws that
the multiplicative process behind Kardashev's
$(1+x)^t$ geometric series assumes.

A linear OLS model and the free-parameter
exponential model fit the data closely and equally
well ($R^2=0.987$ for both; root-mean-square
residuals of $0.48$ and $0.49$~TW, 4\% of the mean
power), and neither is significantly preferred by
the Widely Applicable Information Criterion
($\Delta\mathrm{WAIC}\approx4$, with
$\mathrm{SE}(\Delta\mathrm{WAIC})\approx18$).
Under the two other energy-accounting conventions
the difference stays below two standard errors;
it is largest on the total energy supply, where
the linear form is favoured by 1.8 standard errors
(Section~\ref{sec:limitations}).
However, extrapolation of the linear model to
the Kardashev Type~II threshold yields a
civilisational transition timescale of
$\sim1.1\times10^{5}$ Hubble times, a physical
\emph{reductio ad absurdum}: long before this
date, the Sun will have exhausted its hydrogen
fuel, expanded into a red giant, and likely
engulfed or scorched the Earth. The exponential
model reaches the threshold some 1.5~kyr from
the present, but growth at its fitted rate,
sustained on Earth, would render the planet
uninhabitable through waste heat long before
then \citep{BalbiLingam2025}. Neither of the two
simplest statistically adequate descriptions of
the record therefore yields a physically
coherent path to the energy-scaling Type~II
threshold: this is Kardashev's Conundrum.

The Conundrum has a direct observational
consequence for SETI. Because any planet-bound
trajectory, whatever its growth law, crosses the
waste-heat ceiling of habitability before the
energy-scaling Type~II threshold, Type~II
energy-harvesting technosignatures -- Dyson
spheres in particular -- are unlikely to arise
from a scaled-up terrestrial trajectory; under the
assumptions of Section~\ref{sec:technosignatures},
such structures, if they exist, are the product of
relocating energy dissipation off the home planet.
This offers a candidate explanation for the null
results of existing Dyson-sphere surveys
\citep{Carrigan2009,Wright2014,Suazo2024}, and an
entry in the ``logistic'' category of
Fermi-paradox solutions \citep{Cirkovic2018} that
draws on the waste-heat constraint
\citep{BalbiLingam2025} and converges with the
luminosity-limit constraint \citep{HaqqMisra2025b}
on sustained technological growth.

Conceptually, power alone is a dimensionally
incomplete measure of civilisational advancement.
We have proposed the Kardashev--Sagan--Nakamoto
(KSN) model in response: renormalising the
energy-production ordinate by the annual average
Bitcoin network hashrate $H(t)$, yielding the KSN
state variable $B(t)=P(t)/H(t)$ in units of
$\mathrm{J\,Hash^{-1}}$ -- the KarNak unit. This
renormalisation adds no free parameters and
fulfils the requirement of \citet{Sagan1973} that
the Kardashev scale be sensitive to the
information richness of a civilisation. The KSN
variable spans 14 orders of magnitude over
2009--2024, reflecting both the improving
efficiency of the hardware from the CPU era to
modern ASICs and the growing share of global
energy devoted to the network. Under this
renormalisation the Type~II concept survives the
failure of its energy-scaling realisation on
planet-bound trajectories, as the
approach of the hardware energy per hash to its
Landauer floor. At the present rate of
improvement, the hardware reaches the practical
floor of fast, reliable irreversible logic, nearly
two orders of magnitude above the Landauer floor,
in about three decades
(Section~\ref{sec:timescales}), against the $10^5$
Hubble times of the energy-scaling threshold on
the linear description: this is the sense in which
we call the KSN model a resolution of the
Conundrum.

We further note that \citet{Kardashev1964}
referenced two distinct growth rates: a projected
empirical rate of $3$--$4\%~\mathrm{yr^{-1}}$
(attributed to \citet{Putnam1948}) and a
conservative modelling assumption of
$1\%~\mathrm{yr^{-1}}$ adopted for his Type~II
extrapolation. Our posterior
$r=2.01\%~\mathrm{yr^{-1}}$ falsifies both
simultaneously: the 1\% assumption is too low, and
the 3--4\% projection is too high. The fitted rate
lies between the two values Kardashev referenced,
a nuance lost in a literature that has widely
adopted the 1\% figure.

The KarNak unit, defined on one auditable class of
computation, is proposed as an information-energy
yardstick for the Kardashev scale, and the KSN
framework as a quantitative language for
discussing civilisational longevity in the context
of the Drake equation.

% =====================================================================
\section*{Declaration of generative AI and
AI-assisted technologies in the manuscript
preparation process}

During the preparation of this work the author
used Claude (Anthropic) in order to assist with
the literature search, with the drafting of parts
of the text and with the editing of the
manuscript. The use of the same tool in the
analysis is described in Section~\ref{sec:data}.
After using this tool, the author reviewed and
edited the content as needed and takes full
responsibility for the content of the published
article.

\section*{Declaration of competing interest}

The author declares no competing financial or non-financial interests. The Bitcoin network
hashrate is used solely as a publicly auditable measurement proxy for global proof-of-work
computation; the work advocates no cryptocurrency.

\section*{Acknowledgements}

SG thanks R.\ Gamen from the Argentine Virtual
Observatory (Nuevo Observatorio Virtual Argentino,
NOVA) for the opportunity to serve as Argentina's
representative on the International Virtual
Observatory Alliance (IVOA) Executive Committee
(2015--2024). SG also thanks colleagues at the
Instituto de Astronom\'ia Te\'orica y Experimental
(IATE) and the Observatorio Astron\'omico de
C\'ordoba (OAC) for their support. This work made
use of Astropy, a community-developed core Python
package for Astronomy \citep{Astropy2013}, and the
broader open-source scientific software ecosystem
including SciPy and NumPy. Simulations were run on
the S\'ersic cluster at the IATE. This research
was partially conducted during a period of unpaid
leave from the National Scientific and Technical
Research Council (CONICET), the OAC, and the
National University of C\'ordoba, motivated by
national-government austerity measures that
suspended research funding across Argentine
scientific institutions.

\paragraph{Author contributions (CRediT)}
\textbf{S.\ Gurovich:} Conceptualization,
Methodology, Software, Formal analysis,
Investigation, Writing -- original draft, Writing
-- review \& editing, Visualization, Project
administration.

\section*{Data availability}

The data analysed are public: the global primary
energy series of Our World in Data
\citep{Ritchie2022} and the Bitcoin network data
of Coin Metrics \citep{CoinMetrics2026}. The
series used, and the scripts that reproduce the
figures and the quantitative results of this
paper from them, are available at
\url{https://github.com/sebaguro/KSN_Acta}.

% =====================================================================
% Reference list as produced by BibTeX with the style elsarticle-num-names
% from references.bib (placed here so that no separate file is needed).

\end{document}